# A Generic Deployment Framework for Grid Computing and Distributed Applications

Areski Flissi[1], Philippe Merle[2]

[1] LIFL / CNRS
Université des Sciences et Technologies de Lille
59655 Villeneuve d'Ascq, Lille, France
areski.flissi@lifl.fr

[2] INRIA Futurs / LIFL
Université des Sciences et Technologies de Lille
59655 Villeneuve d'Ascq, Lille, France
philippe.merle@inria.fr

**Abstract.** Deployment of distributed applications on large systems, and especially on grid infrastructures, becomes a more and more complex task. Grid users spend a lot of time to prepare, install and configure middleware and application binaries on nodes, and eventually start their applications. The problem is that the deployment process is composed of many heterogeneous tasks that have to be orchestrated in a specific correct order. As a consequence, the automatization of the deployment process is currently very difficult to reach. To address this problem, we propose in this paper a generic deployment framework allowing to automatize the execution of heterogeneous tasks composing the whole deployment process. Our approach is based on a reification as software components of all required deployment mechanisms or existing tools. Grid users only have to describe the configuration to deploy in a simple natural language instead of programming or scripting how the deployment process is executed. As a toy example, this framework is used to deploy CORBA component-based applications and OpenCCM middleware on one thousand nodes of the French Grid5000 infrastructure.

**Keywords:** Grid computing, middleware, distributed applications, deployment, software components.

## 1 Introduction

Thanks to the availability of some grid infrastructures as Grid5000 [1] –the French grid infrastructure for research– software deployment tends to become a major research topic for software engineering and distributed systems communities. Deployment can be defined as a set of tasks to realize such as software installation/uninstallation on remote nodes and activation/deactivation of software instances. These deployment tasks have to be orchestrated in a defined order. For instance, software activation cannot be done until it has been downloaded, installed and configured. Many works try to address, at different levels, the problem of software deployment on large distributed systems. On one hand, in the field of middleware, most of component-based platforms offer deployment facilities allowing to deploy component-based applications. For example, the OSGi technology [2], which is dedicated to services executed on gateways, provides mechanisms for remotely deploying and managing OSGi bundles representing end-user services. JOnAS [3] or OpenCCM [4] platforms also provide the same kind of features to respectively deploy J2EE and CCM applications. But, these platforms do not address the problem of the middleware deployment itself. Concerning grids, ProActive middleware [5] addresses the deployment of servers and applications but built on top of ProActive only.

On the other hand, many *adhoc* solutions for deploying software, as file transfer protocols to download and install software on remote nodes (*e.g.*, FTP, HTTP, scp, rsync) or remote access protocols to launch processes (*e.g.*, SSH, telnet, rlogin), exist and are often used. In addition, grid infrastructures themselves provide sometimes end-users tools to deploy applications. For instance, Grid5000 provides the OAR tool [6] for reservation of nodes and Kadeploy [7] for deployment of customized operating system images.

Thus, software deployment on large infrastructures (grids are composed of thousands nodes) becomes a more and more complex task: Grid users spend a lot of time (from hours to days!) to prepare, configure nodes, install middleware and software and finally start their experiences. In support of this view, let's just take an example about deployment of CCM applications on the Grid5000 infrastructure. At first, users must connect to the frontal node (using *ssh*) of one cluster of the grid, check for available nodes, submit a reservation request using OAR, and get the list of allocated nodes. Then, the CCM application servers that will host applicative components have to be installed on each node, plus all required libraries and external stuffs (*e.g.*, a *Java Runtime Environment* or a CORBA runtime). To achieve that, users will probably use classical techniques to download, unzip and install binaries on remote nodes or can use Kadeploy to deploy a specific image previously built. Unfortunately, this second step has to be repeated for each experience because Grid5000 is a shared infrastructure, which means nodes are reinitialized by next users. After that, component servers have to be started on each node, but also useful middleware services. However, some dependencies exist between these processes. For example, the CORBA Naming Service must be started on a node before all CCM component servers. At this stage, the execution support is finally ready for the deployment of the application. This is true here (but not always) as the OpenCCM platform provides a distributed deployment infrastructure named DCI [8] that manages the "high-level" deployment tasks, –*i.e.*, installation and instantiation of component binaries on nodes, binding of components and finally activation of component instances.

Our contribution in this paper is about a generic deployment framework dedicated to large systems as grid infrastructures. Our approach consists in reifying existing mechanisms and tools involved in the deployment process as Fractal components [9]. These components are composed together which allows to assemble heterogeneous mechanisms for an automatized execution. Software to deploy, whatever the granularity is, and physical infrastructure are also reified as components. Bindings between components represent dependencies. Finally, the assembly of components represents the configuration of the system to deploy.

The reminder of the paper is organized as follows. Section 2 describes the key principles of our framework and the generic deployment components. Section 3 explains how to write software personalities for deployment of specific software and gives, as an illustration, a concrete example about the deployment of OpenCCM middleware on Grid5000. In Section 4, related works are discussed, and finally, Section 5 concludes this paper and gives some future works.

## 2   A Generic Deployment Framework

### 2.1   Overview of the approach

Our approach is based on a generic deployment framework allowing to build, using personalities, the appropriate tool to automatically orchestrate the deployment process of software, regardless of the technology or the granularity (from middleware platforms, application servers to applications, services, or even objects), as illustrated in Figure 1.

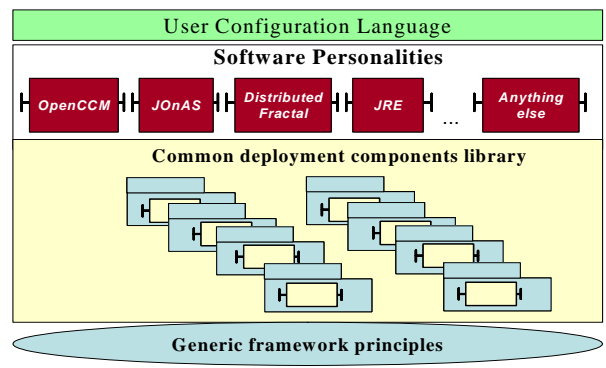

**Fig. 1.** Overview of the generic deployment framework

The deployment framework first relies on some key principles, as "everything is reified as components", provides a common library of deployment components, and finally personalities written once by developers for specific technology or software. The main advantage for users of such a tool is to allow to just describe configurations –*i.e.*, nodes of the targeted infrastructure and services[1] to deploy on them– instead of programming the way deployment has to be executed and without taking care of the orchestration of deployment tasks. As a consequence, automatization of heterogeneous activities becomes possible thanks to our framework.

---

[1] The word "*services*" is used in this paper as a generic term to reference any kind of "software": components, objects, scripts, programs, middleware services, application servers, processes, etc.

### 2.2 Everything reified as components

First of all, the main requirement of our work was not to re-invent the wheel but to rely on existing mechanisms. In other terms, our goal is to provide, through a generic framework, a simple way to reuse and especially compose existing heterogeneous tools for automatic deployment of software on very large systems as grids. The first key principle of our approach is that everything is reified as software components [10], or more precisely Fractal components. "Everything" here includes all the elements that are manipulated or involved in the deployment process:
- Physical hosts, such as Grid5000 nodes
- Remote access protocols, such as *SSH*, *Telnet*, etc.
- File transfer protocols, such as *FTP*, *HTTP*, *scp*, *rsync*
- Shells, such as *SH*, *CSH* or *Windows Command Shell*
- Shell variables and commands
- Software and tools, such as *Java Runtime Environment*, *OAR* or *Kadeploy*
- Middleware services, such as a *CORBA Naming Service*
- Application servers, such as *OpenCCM*, etc.

The second principle is to compose these components in order to obtain an assemby (or composite) which symbolizes software to deploy. In other words, the execution of this composite corresponds to the execution of the deployment process. Bindings between components represent dependencies. For example, the *Shell* component reifies a "real" shell and provides, through a server interface, a method to execute shell commands. For that, it needs a remote access protocol to send the command to a remote host. So, *Shell* component is bound to a *Protocol* component, itself bound to other components for getting hostname, port and user access information.

Finally, the third principle of our approach, is that orchestration of components contained in the global composite is done by the framework, not by the user. To allow that, every component contained in the composite provides a server interface named `Deployment` that offers the generic methods `install()`, `start()`, `stop()` and `uninstall()`. Components are executed according to the state diagram logic (install if is uninstalled then start; stop if is started then uninstall) illustrated in Figure 2 and in an order that is determined by bindings between components. For instance, in the case of the JRE software, that is used by many other software as OpenCCM or JOnAS, we can draw different scenarios depending on the fact the JRE is already installed or not on the node. If not, the JRE has to be installed before, whereas if it is already installed, it has just to be started –*i.e.* set the `JAVA_HOME` shell variable and add *java* binary to the `PATH` environment variable using the *deployment components*.

This process is very similar to the orchestration framework of tasks we defined in [11]. The choice of Fractal is well suited here since it is a hierarchic component model supporting sharing of component instances and component introspection operations. The *specific action* mentionned in Figure 2 represents a particular "business" operation provided by another server interface of the component (*e.g.,* the `send()` method of the *Protocol* component) and used by another component (*e.g.,* the *Shell* component).

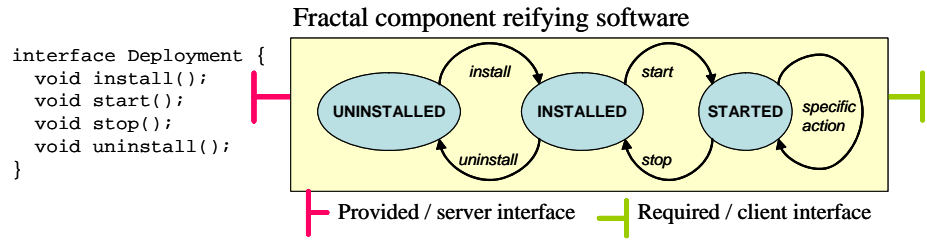

**Fig. 2.** Generic state diagram used for orchestration of components

### 2.3 A common reusable library of deployment components

We have developed a set of reusable deployment components, that, relying on our deployment framework and combined to specific components (software personalities) can be composed together by users to describe a configuration and automatically deploy it on the grid. These components provide common server interfaces that are used during the deployment process. These main interfaces are described in Table 1. Many implementations of these interfaces are possible. Thus, the implementation of the *Protocol* interface can be based on *SSH* or *Telnet* whereas the implementation of *Shell* can be *SH*, *CSH* or *Windows Command*. What should be emphasized again here is that our components do not re-implement things. For example, we developed one *Telnet* implementation of the *Protocol* component that uses a free open source external library (named *JTelnet*) and one implementation of the *SSH* component that uses external OpenSSH process.

**Table 1.** Main component interfaces of our generic deployment framework

| Interface | Main offered methods and description |
| --- | --- |
| Hostname | `getInternetHostname()`: Return or (dynamically) compute an Internet hostname (or IP address). |
| Port | `getInternetPort()`: Return an Internet port for protocols (*e.g.,* 22 for *SSH*). |
| User | Provides methods for getting user access information (login, password, RSA private key, etc.). |
| Protocol | `send(args)`: Send commands to a remote host. |
| Shell | `execute(args)`: Execute shell commands. |

On top of these fundamental deployment components, we wrote some other reusable components such as JRE. JRE can be defined as a *service* to "start" in user's configuration or can be used by other service components (*e.g.,* OpenCCM or JOnAS need a JRE). As explained in Section 2.2 (all *services* are reified as components that implement the *Deployment* interface), JRE implements install, start, stop and uninstall methods. For that, it is bound to fundamental deployment components (such as *Shell*, *Variable*, etc.) and others like a *File Transfer* component (for installation of the JRE archive from the SUN web site!). Figure 3 illustrates an example of assembly for this particular useful component.

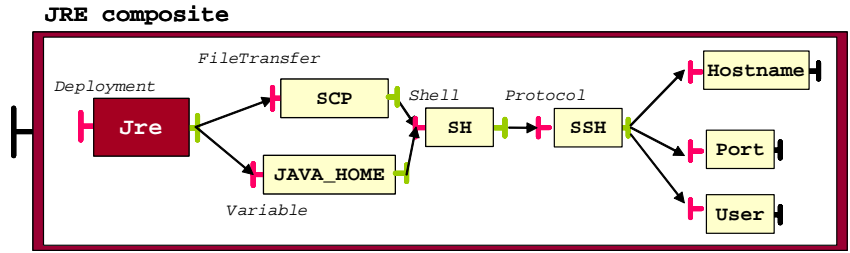

**Fig. 3.** An example of assembly for JRE component

Assemblies and complex composites are described once by developers via Fractal *Architecture Description Language* (ADL) files. This allows users to ignore bindings (dependencies) between components and provides a high level of abstraction of the deployment process: deployment mechanisms, dependencies between software or orchestration complexity are hidden for users and managed by the framework.

## 3 Illustration: Automatic Deployment of OpenCCM on Grid5000

This section illustrates a concrete example of the use of our deployment framework to deploy the OpenCCM middleware on Grid5000.

### 3.1 Write personalities once, work everytime

For developers, writing personalities mainly consists in writing Fractal ADL files describing composites of components, and in some cases, specific components implementing specific business code. At first, in the case of OpenCCM, we wrote few composites and primitive components that reify OpenCCM entities: OpenCCM (as the software), NS (the CORBA Naming Service), DCI (*OpenCCM DCI Manager* that manages distributed deployment of CORBA components) or *DCI_NODE* (servers that host CORBA components). These composites are a composition of our generic deployment components and of components they depend on, as JRE. Secondly, in the purpose of deployment on Grid5000 infrastructure and according to the same principles used for OpenCCM components, we wrote components for OARGrid (reservation of nodes on many clusters of Grid5000) and Kadeploy (deployment of customized images) tools.

We have finally written a specific implementation of the *Hostname* interface allowing to provide a dynamic hostname used by the *Protocol* component. Indeed, hostnames of nodes allocated by OARGrid aren't known in advance. Moreover, as we will see in Section 3.2, this allows users to describe configurations of the physical infrastructure simply by "telling" that hostnames are dynamically computed (this point is very useful for Grid5000 as currently grid users' applications have to manage themselves a dynamic list of allocated nodes).

### 3.2 Write user configurations to deploy

Let's now see what users finally have to write in order to automatically initiate a complex heterogeneous deployment process. As we have seen in previous sections, the main goal is to allow users to just describe in a simple natural language their configurations (the services to deploy and the targeted nodes) without taking care of how things are done. An example of a simple configuration is given below.

```
MyDeployment = OpenCCM.Deployment {
  nodes = {
        hostname = DynamicHost(~/nodelist)
        apply FOR(i,0,500) {
           node-%{i} = Grid5000_NODE {
                  hostname = nodes/hostname
                  protocol = SSH
                  port = Port(22)
                  shell = SH
                  user = User(aflissi, ,~/.ssh/id_rsa.pub)
                  jre = Jre(/opt/java/jdk-1.5.0_05)
                  openccm = OpenCCM(/opt/OpenCCM-0.9,/opt/CORBA/JacORB-2.2)
  } } }
  services = {
        ns = OpenCCM.NameService {
              node = nodes/node-0 }
        dci = OpenCCM.DCIManager(MyDCI) {
             ns = services/ns
             node = nodes/node-0 }
        servers = ParallelRunner {
             apply FOR(i,1,500) {
                 server-%{i}= OpenCCM.DCI_NODE(NM_%{i}){
                          dci = services/dci
                          node = nodes/node-%{i}

} } } } }
```

This piece of code uses a simple *Domain Specific Language* (DSL) that is composed of two parts: the first one (*nodes {...}*) describes configuration of the targeted nodes and the second one (*services {...}*) specifies services to start and where to start them. Here the user simply describes that: 1- he/she uses Grid5000 nodes with specific configurations for protocol (SSH), shell (SH), JRE, OpenCCM; hostnames are dynamically computed, and 2- he/she wants to start an OpenCCM server (DCI_NODE) on each node (*node-1* to *node-500*) and the *CORBA Naming Service* (NS) on *node-0*. Services are by default sequentially executed (*ns* first, then *dci*, then *servers*) but user can specify particular behavior using tag such as *ParallelRunner* for parallel deployment of servers. Other examples of *services* can be specified in the configuration: OARGrid, Kadeploy or anything else. If a reservation task is needed, the user simply has to add it in the configuration thanks to the following lines:

```
  nodes = {   oar_server = Grid5000_NODE {
                  hostname = StaticHost(oar.lille.grid5000.fr)
  } }
  services = { reservation =
               OARGrid(gdx=300|azur=100|grillon=50|lille=50,~/nodelist) {
                 node = nodes/oar_server

  } }
```

From user's configuration, a Fractal ADL file is generated and interpreted by the Fractal ADL tool to instantiate the global composite that abstracts the system to deploy. The generated Fractal ADL file contains definitions of both generic deployment components and software personalities. The deployment framework, relying on existing infrastructure, is then in charge of the orchestration of tasks, the management of software dependencies, or the execution of specified policies –i.e., sequential or parallel deployment of services. Figure 4 illustrates the deployment of our example on the Grid5000 infrastructure. Components contained in the global composite are executed according to a specific order computed by our orchestration logic. In this example, OARGrid is started first since it has no dependencies (no required interfaces): inside the OARGrid composite, fundamental deployment components (shell, protocol, hostname, port, user, etc.) are started to send the *oarsub* request –*i.e.*, submit a reservation– to a remote server of Grid5000 (here *oar.lille.grid5000.fr*). Then, the OpenCCM component is executed, which means: install OpenCCM middleware using a *File Transfer* component to download binary from OpenCCM web site, and send commands on each remote node to set and configure OpenCCCM shell variables. Next, the NS component and at least the *DCI_NODE* components (CCM application servers) are started.

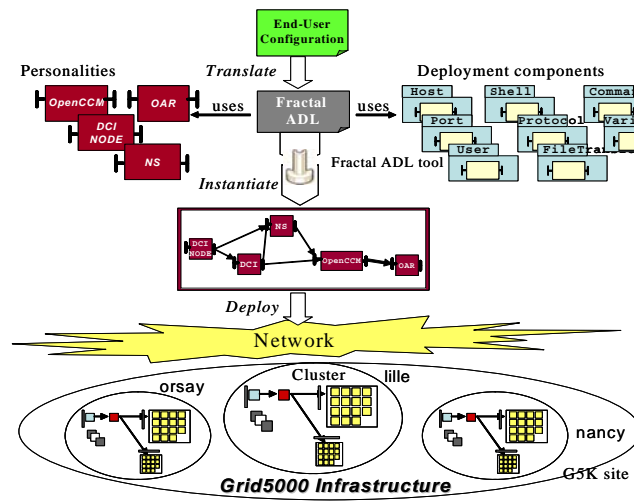

**Fig. 4.** Automatic deployment of the OpenCCM middleware on Grid5000

### 3.3 Evaluation and Performances

Deployment of OpenCCM middleware up to one thousand nodes of Grid5000 have been successfully tested using our framework. For this case, effective deployment time (without the overhead introduced by the framework when parsing the user's configuration file and that is due to the Fractal ADL tool) takes less than 300 seconds, with a parallel execution. Indeed, deployment process is composed of two main steps: 1) loading of configuration and instantiation of components by the Fractal ADL tool, and 2) execution of the deployment process itself. What's more, effective deployment time is approximatively growing linearly with the number of nodes. All experimental results are showed in Figure 5. Grid5000 "virtual" nodes means here that many application servers may have been started on a same physical node.

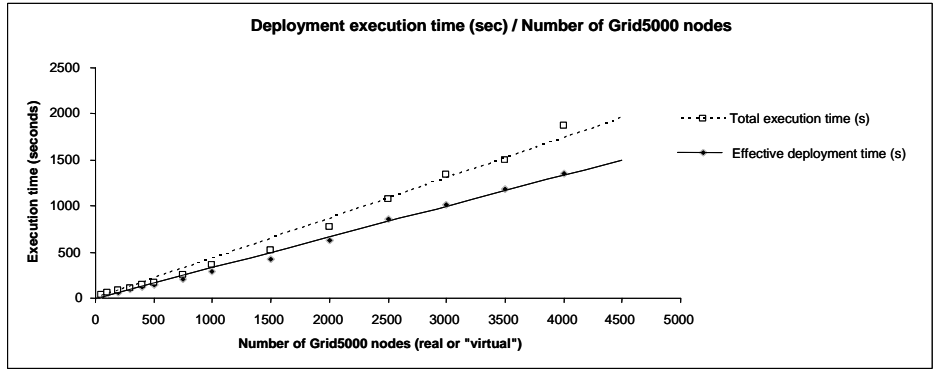

**Fig. 5.** Execution time of the deployment process of OpenCCM middleware on Grid5000

## 4 Related Works

Many works are done on the deployment of distributed applications on large scale systems. Concerning application servers and gateways, technological platforms such as OSG*i* or J2EE address the problem of automatic deployment of applications. But, OSGi does not address deployment of distributed applications. Concerning both OSGi and J2EE, nothing is said about deployment of the middleware platforms themselves on gateways/servers. This task has to be done manually by administrators. OpenCCM encountered the same lack as it can only deploy CCM-based applications. Other works in the *Autonomic Computing* domain are based on the software component paradigm. JADE [12] uses Fractal components to reify the software for deployment and dynamic reconfiguration of J2EE systems. Nevertheless, JADE infrastructure must have been installed before on targeted platforms.

In the field of Grid Computing [13], ProActive allows the deployment of both ProActive-based servers and applications but is limited to this technology. *S. Lacour* proposes in [14] a software architecture for the automatization of the deployment process of distributed applications executing on grid infrastructures. Our goal is the same but the approach is different. Main key points are that, in one hand we rely on existing infrastructures, tools or paradigms for applications and middleware, regardless of the granularity, and, in the other hand we simplify the description of complex deployment's configurations (in a single file if needed).

## 5 Conclusion and Future Works

Composing heterogeneous mechanisms in order to automatize the deployment process of software executed on large scale infrastructures such as grids is the next challenge. Our approach is based on the *software component* paradigm. We have defined and implemented in Fractal a generic deployment framework and fundamental components that can be used to build specific deployment tools. Users only have to describe their configurations in a simple language and use the framework to instantiate the deployment. To validate the approach, we have successfully deployed the OpenCCM middleware on one thousand nodes of Grid5000.

Our future work will go in many directions. We first plan to implement personalities for other technologies such as J2EE or OSGi servers. Secondly, in order to optimize very large scale deployment, we will enhance the framework with the monitoring and control of local resources (sockets, threads, etc.) by distributing it. Finally, we would like to add *dynamic adaptation* aspects in our framework. For instance, as for computation of Grid5000 hostnames, composites reifying services could be automatically assembled or reconfigured at runtime thanks to a dynamic discovery of available protocols, ports or shells for each node.